\documentclass[%
reprint,
superscriptaddress,
amsmath,amssymb,
aps,
prb,
floatfix,
]{revtex4-2}

\usepackage{graphicx}
\usepackage{dcolumn}
\usepackage{booktabs}
\usepackage{bm}
\usepackage{braket}
\usepackage[colorlinks]{hyperref}
\usepackage{color}
\usepackage[usenames,dvipsnames]{xcolor}
\usepackage{comment}
\usepackage[export]{adjustbox}
\usepackage{siunitx}
\usepackage{fouriernc}

\emergencystretch=\maxdimen
\newcommand{\be}{\begin{equation}}
\newcommand{\ee}{\end{equation}}

\begin{document}
	
	\small
	
	\title{Kinetic inductance in superconducting CoSi\textsubscript{2} coplanar microwave transmission lines }
	
	\author{Ekaterina Mukhanova}
	\affiliation{Low Temperature Laboratory, Department of Applied Physics, Aalto University, P.O. Box 15100, FI-00076 Espoo, Finland}
	
	\affiliation{QTF Centre of Excellence, Department of Applied Physics, Aalto University, P.O. Box 15100, FI-00076 Aalto, Finland}%

    \author{Weijun Zeng}
	\affiliation{Low Temperature Laboratory, Department of Applied Physics, Aalto University, P.O. Box 15100, FI-00076 Espoo, Finland}

	\affiliation{QTF Centre of Excellence, Department of Applied Physics, Aalto University, P.O. Box 15100, FI-00076 Aalto, Finland}
 
    \author{Elica Anne Heredia}
    \affiliation{International College of Semiconductor Technology, National Yang Ming Chiao Tung University, Hsinchu 30010, Taiwan}

	\author{Chun-Wei Wu} 
    \affiliation{Department of Electrophysics, National Yang Ming Chiao Tung University, Hsinchu 30010, Taiwan}
 
	\author{Ilari Lilja}
	\affiliation{Low Temperature Laboratory, Department of Applied Physics, Aalto University, P.O. Box 15100, FI-00076 Espoo, Finland}
	
	\affiliation{QTF Centre of Excellence, Department of Applied Physics, Aalto University, P.O. Box 15100, FI-00076 Aalto, Finland}

    \author{Juhn-Jong Lin}
	\affiliation{Department of Electrophysics, National Yang Ming Chiao Tung University,
Hsinchu 30010, Taiwan}

 \author{Sheng-Shiuan Yeh}  \email{ssyeh@nycu.edu.tw}
	
	\affiliation{International College of Semiconductor
Technology, National Yang Ming Chiao Tung University, Hsinchu 30010, Taiwan}

	\affiliation{Center for Emergent Functional Matter Science, National Yang Ming Chiao Tung University, Hsinchu 30010, Taiwan}

	\author{Pertti Hakonen}  \email{pertti.hakonen@aalto.fi}
	\affiliation{Low Temperature Laboratory, Department of Applied Physics, Aalto University, P.O. Box 15100, FI-00076 Espoo, Finland}
	
	\affiliation{QTF Centre of Excellence, Department of Applied Physics, Aalto University, P.O. Box 15100, FI-00076 Aalto, Finland}

	\date{\today}
	
	\begin{abstract}

We have looked into cobalt disilicide (CoSi\textsubscript{2}) as a potential building block for superconducting quantum circuits. In order to achieve this, we annealed a thin layer of Co to create \qtyrange[range-phrase = --,range-units = single]{10}{105}{nm} thick microwave cavities from CoSi\textsubscript{2} embedded in the silicon substrate. The cavity properties were measured as a function of temperature and power. In films measuring 10 and 25~nm, we find a significant kinetic inductance $L_\mathrm{K}$ with a non-BCS power-law variation $\delta L_\mathrm{K} \propto T^{4.3 \pm 0.2}$ at low temperatures. The quality factor of the studied microwave resonances increased almost linearly with thickness, with two-level systems having very little effect. The power dependence of kinetic inductance was analyzed in terms of heat flow due to electron-phonon coupling, which was found stronger than estimated for heat relaxation by regular quasiparticles.
  
	\end{abstract}
	
	\maketitle
	

\section{Introduction}

 Microwave cavities are essential building blocks for quantum circuits and circuit quantum electrodynamics \cite{Blais2021}. Their low frequency noise properties \cite{Falci2023} and dissipation at GHz frequencies \cite{Wang2020} will influence, for example, qubit decoherence, which in turn causes errors in qubit gates as well as degrades the single-shot read out efficiency. In addition to the high quality of a fabrication material \cite{deLeon2021}, the material's compatibility for circuit fabrication is an essential asset for the utilization of very large scale integration technology for superconducting circuits \cite{Tolpygo2016}.  
 
 In this work, we investigate the potential of cobalt disilicide (CoSi\textsubscript{2}) as a high-quality material for quantum circuits. Silides, for example vanadium silicide \cite{Vethaak2022}, are known to be promising materials for silicon-based superconducting field effect transistors provided that the problem of Schottky barriers is overcomed \cite{Schwarz2023}. A cobalt disilicide conductor is a low-temperature superconductor that can be obtained straightforwardly by annealing a stripe of cobalt laid on top of the silicon substrate. The induced chemical reaction is self-limiting and it results in a high-quality, nearly epitaxial CoSi\textsubscript{2} film with excellent lattice match to silicon. Owing to excellent lattice match, the amount of disorder is small at the CoSi\textsubscript{2}/Si interface which promises low $1/f$ noise in this material \cite{Chiu2017}.

 Thin CoSi\textsubscript{2} films possess a substantial kinetic inductance. This inductance is related to normal state resistance $R_\mathrm{n}$ which is known to display low frequency resistance fluctuations $(\delta R_\mathrm{n})^2 \propto R_\mathrm{n}^2$ with $1/f$ spectrum \cite{Falci2023}. The $1/f$ resistance noise of 105-nm-thick CoSi\textsubscript{2} films has been investigated in Ref.~\onlinecite{Chiu2017}, and the noise was found to be quite small. One of the questions that we address is whether large kinetic inductance can be achieved in CoSi\textsubscript{2} without enhanced dissipation at microwave frequencies caused typically by the presence of two-level tunneling states (TLS) \cite{Lisenfeld2019,Wang2020}. A high-quality superconductor with large kinetic inductance would facilitate, for example, wideband traveling wave parametric amplifiers \cite{Eom2012,Chaudhuri2017}.

The kinetic inductance for a superconducting wire with the length $L_\mathrm{line}$ is given near zero temperature by \cite{Bezryadin2013} 
\be
L_\mathrm{K}=\frac{2}{3\sqrt{3}}\frac{L_\mathrm{line}}{\xi}\frac{\hbar}{2e I_\mathrm{c}}=\frac{\hbar R_\mathrm{n}}{\pi\Delta}.
\label{LK}
\ee
In the latter form, $\Delta$ denotes the superconducting gap and $R_\mathrm{n}$ denotes the total normal-state resistance of the sample, and we have employed the fact that, according to the Ambegaokar-Baratoff-type of relation, the product $R_\mathrm{n}(\xi)I_\mathrm{c} =\frac{\pi}{3\sqrt{3}} \Delta/e$ \cite{Tinkham2002}, where $R_\mathrm{n}(\xi)$ denotes the resistance of a film section of coherence length $\xi$; 
 for CoSi\textsubscript{2}, the superconducting coherence length at zero temperature $\xi(0)=90$~nm \cite{Chiu2021}. Consequently, the kinetic inductance in thin CoSi\textsubscript{2} films is enhanced both by increased film resistance and the reduced transition temperature. At finite temperatures there is a correction $\propto \tanh(\Delta/k_\mathrm{B} T)^{-1} $ \cite{Maki1963, Annunziata2010}, which is insignificant in our analysis and has been omitted.

\section{Sample fabrication and characteristics}
Our samples use a regular notch-type design with five cavities in parallel as shown in Fig.~\ref{fig:layout}a. A cobalt film was deposited via thermal evaporation onto a surface area of an undoped Si(100) substrate defined by photolithography. Before the deposition, the native SiO\textsubscript{2} on the substrate surface was removed by aqueous hydrogen fluoride (HF) solution. The deposited Co film was annealed in a vacuum with a pressure of \qty{\sim 1e-6}{Torr} for silicidation. During the thermal annealing process, Co atoms act as predominant moving species, diffusing downwards and reacting with Si atoms to form the epitaxial CoSi\textsubscript{2} film \cite{Chen2004}. The silicide thickness ratio for CoSi\textsubscript{2}/Co is \num{\approx 3.5} \cite{Chen2004}, i.e., for a Co film with thickness $t_{\rm Co}$, after silicidation, the thickness of CoSi\textsubscript{2} is \num{\approx 3.5}$t_{\rm Co}$. A cross-sectional view of our device structure is given in Fig.~\ref{fig:layout}b.

In this work, we deposited Co films with thicknesses of 3.5, 7, and 30~nm, and the expected thicknesses for CoSi\textsubscript{2} films are \num{\approx 10}, \num{\approx 25}, and \qty{\approx 105}{nm}, respectively. The annealing conditions for 3.5, 7, and 30-nm-thick Co films are \qty{550}{\degreeCelsius} (30~min) followed by \qty{600}{\degreeCelsius} (30~min), \qty{600}{\degreeCelsius} (30~min) followed by \qty{700}{\degreeCelsius} (30~min), and \qty{700}{\degreeCelsius} (1~hr) followed by \qty{800}{\degreeCelsius} (1~hr), respectively. Fig.~\ref{fig:layout}c displays an AFM image of the surface roughness in our 25-nm-thick samples. The measured root mean square (rms) roughness of CoSi\textsubscript{2} surface (excluding the edges with extra spikes) amounts to  1.56, 1.73, and 1.78\, nm for the samples with the thickness of 10, 25, and 105\, nm, respectively.  The superconducting temperatures $T_\mathrm{c}$ are determined by measuring the temperature dependence of resistance at low $T$ using a \textsuperscript{3}He cryostat. The $T_\mathrm{c}$ is 1.37, 1.31, and 1.35~K for the CoSi\textsubscript{2} films with thicknesses of $d=10$, 25, and 105~nm, respectively. The normal-state sheet resistances $R_\mathrm{s}$ at 1.5~K are 7.1, 2.4, and \qty{0.24}{\ohm}, respectively.

Nominally, the cavities have a characteristic impedance of \qty{50}{\ohm} but this is valid only as long as the kinetic inductance can be neglected in the design ($d \gtrsim 100$~nm). We chose 10 micron width for the center conductor on a high-purity silicon substrate, for which a 6 micron gap between the center conductor and the ground plane corresponds to a \qty{50}{\ohm} transmission line. Afterward, using the measured kinetic inductance, we may state the actual estimates for characteristic impedance as 88, 65, and \qty{50} {\ohm} for thicknesses 10, 25, and 105~nm, respectively. We note that the ground plane conductor is made of the same material, which means that the ground plane does contain significant kinetic inductance for our 10~nm and 25~nm resonators, but this was neglected in our analysis.

 
\section{Experimental methods}

Our experiments were mostly performed on a Bluefors LD400 dilution refrigerator with a base temperature of 10~mK. The refrigerator has a \qtyrange[range-phrase = --,range-units = single]{4}{8}{GHz} microwave measurement setup which, in addition to a Low Noise Factory HEMT amplifier, encompasses a Josephson traveling wave parametric amplifier (JTWPA) with nearly quantum limited operation for \qtyrange[range-phrase = --,range-units = single]{4}{7}{GHz} frequency \cite{Perelshtein2022}. A diplexer was employed to combine the signal and pumping frequency in the JTWPA. Two \qtyrange[range-phrase = --,range-units = single]{4}{8}{GHz} circulators were installed between sample and JTWPA while two \qtyrange[range-phrase = --,range-units = single]{4}{12}{GHz} circulators were employed at JTWPA output to eliminate back-propagating pump signal. For further details of the system, we refer to Ref.~\onlinecite{Noise}.

The calibration of the measurement power was done employing the known system noise temperature $T_\mathrm{N} = 5$~K for the HEMT amplifier. This calibration yielded 79.6~dB of attenuation for the ingoing microwave signal around 6~GHz (frequency range of 25-nm-thick samples). This attenuation was subtracted from the generator power, which in our experiments was limited to +15~dBm. The calibration was done similarly for the experiments on 10~nm ($\simeq$4~GHz ) and 105~nm devices ($\simeq$7~GHz). 

\begin{center}
    \begin{figure}[tb]
        \centering
        \includegraphics[width=\linewidth]{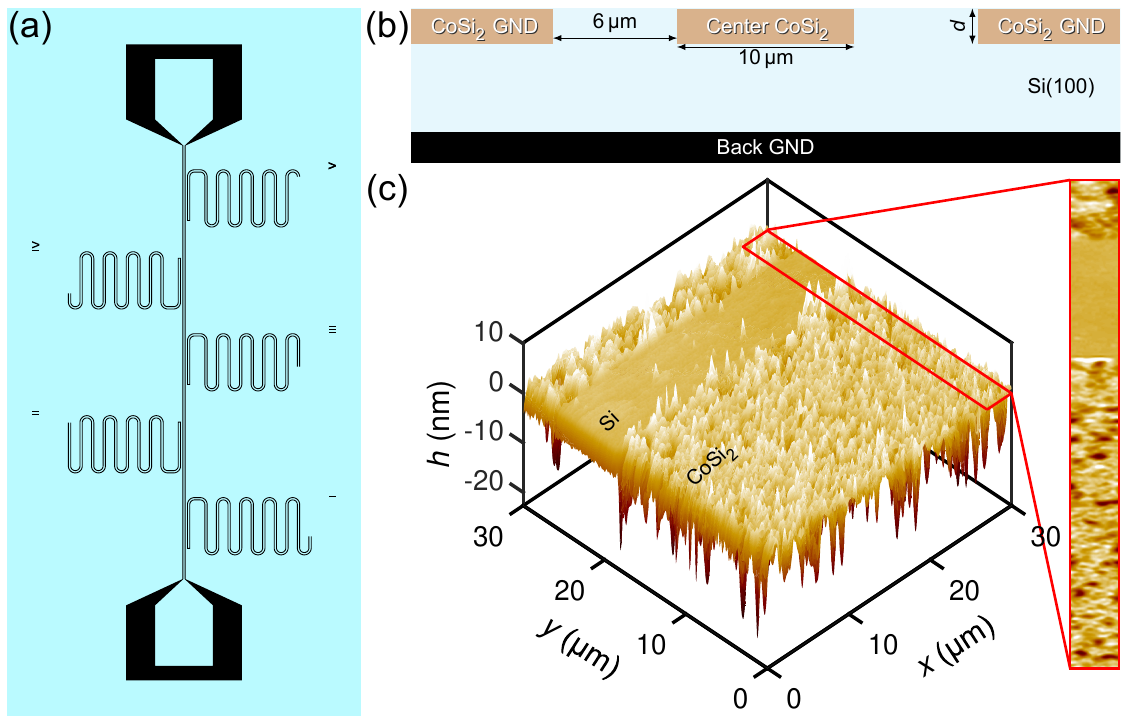}
            \caption{a) Notch-type resonator layout with five cavities with slightly different lengths. b) The cross-sectional view illustrates the diffusion of cobalt into the silicon crystal and the ensuing sample structure. c) AFM image of the surface structure of our 25-nm-thick cavity: Si and CoSi\textsubscript{2} are marked in the image. The CoSi\textsubscript{2} films have dips at the surface, which may act as trapping sites for flux lines in the sample.}
        \label{fig:layout}
    \end{figure}
\end{center}

Our notch-type samples have five coplanar, quarter-wave microwave cavities connected capacitively to the measurement line with a separation of approximately 0.6~mm (see inset of Fig.~\ref{fig:layout}. The lengths of the $\lambda/4$ cavities vary over \qtyrange[range-phrase = --,range-units = single]{3.86}{4.36}{mm}, which corresponds to the design frequency range of $f_{\mathrm{dgn}}=$~\qtyrange[range-phrase = --,range-units = single]{6.83}{7.73}{GHz} without kinetic inductance. The estimation of kinetic inductance from the experimental results was obtained using the basic equivalent model of $\lambda/4$ cavity with resonant frequency $2\pi f_\mathrm{m}=1/\sqrt{C(L_{\mathrm{geo}}+L_\mathrm{K})}$ where $C$ is the total capacitance, $L_{\mathrm{geo}}$ is the total geometric inductance, and $L_\mathrm{K}$ is the corresponding kinetic inductance of the coplanar line.

 The scattering parameter for transmission, $S_{21}(\omega)$, for a notch type resonator is given by \cite{Gao2008,Khalil2012,Weides2015}
 \begin{equation}
S_{21}(\omega) = a\mathrm{e}^{\mathrm{i}\alpha}\mathrm{e}^{-\mathrm{i}\omega\tau}\left( 1 - \frac{\left(Q_\mathrm{load}/|Q_\mathrm{cpl}|\right)\mathrm{e}^{\mathrm{i}\varphi} }{ 1 + 2\mathrm{i}Q_\mathrm{load}\left(\omega/\omega_\mathrm{r} - 1\right) } \right)
\label{eq:notch}
\end{equation}
where $a\mathrm{e}^{\mathrm{i}\alpha}\mathrm{e}^{-\mathrm{i}\omega\tau}$ quantifies the environment with amplitude $a$, phase shift $\alpha$ and electronic delay $\tau$, $Q_\mathrm{load}$ the loaded quality factor, $|Q_\mathrm{cpl}|$ the absolute value of the coupling quality factor, $\varphi$ quantifies impedance mismatch, $\omega$ the probing frequency and $\omega_\mathrm{r}$ the resonance frequency. Fitting \cite{resonatortools} of Eq.~(\ref{eq:notch}) to the measured data yields $Q_{\mathrm{cpl}}$ and $Q_{\mathrm{load}}$, which then gives the internal $Q$-factor using
\be
\frac{1}{Q_{\mathrm{int}}}=\frac{1}{Q_{\mathrm{load}}}-\frac{1}{Q_{\mathrm{cpl}}}.
\ee
In our data on 10~nm and 105~nm samples, the resistive and reactive responses are somewhat mixed, which requires careful analysis using Eq.~(\ref{eq:notch}). In the experiments, the magnitude of the transmission signal is well calibrated, so that the parameter $a=1$ with an error \num{<1}\%. Also, the phase of the VNA probing signal is quite well calibrated, and $\omega\tau \ll 1$ in most of our response curve fits, except some fits at large measurement powers (data in the emerging non-linear Duffing regime).

The internal quality factor is expected to be governed by dissipation coming from quasiparticle resistance (qp), two-level tunneling states (TLS), and vortex flow (FF) \cite{Wang2020}:
\be
\frac{1}{Q_{\mathrm{int}}}=\frac{1}{Q_{\mathrm{qp}}}+\frac{1}{Q_{\mathrm{TLS}}}+\frac{1}{Q_{\mathrm{FF}}}. \label{eq:Q}
\ee
Here $Q_{\mathrm{qp}}=R_{\mathrm{qp}}/\omega L$ with $R_{\mathrm{qp}}=R_\mathrm{n}\exp{(\Delta/k_{\mathrm{B}}T})$ and $L$ the total inductance, and $Q_{\mathrm{TLS}}$ is typically a function of average photon occupation number $\bar{n}$ in the cavity, i.e. $Q_{\mathrm{TLS}}(\bar{n})$ with a tendency to lead to an increase in $Q$ with increasing power owing to saturation and relaxation behavior of the TLSs. The last term in Eq.~(\ref{eq:Q}) describes dissipation due to vortex motion under the constraints by defects/pinning centers in the sample.

	
\section{Experimental results}
\begin{figure*}[!tb]
     \centering
     \includegraphics[width=0.9\linewidth]{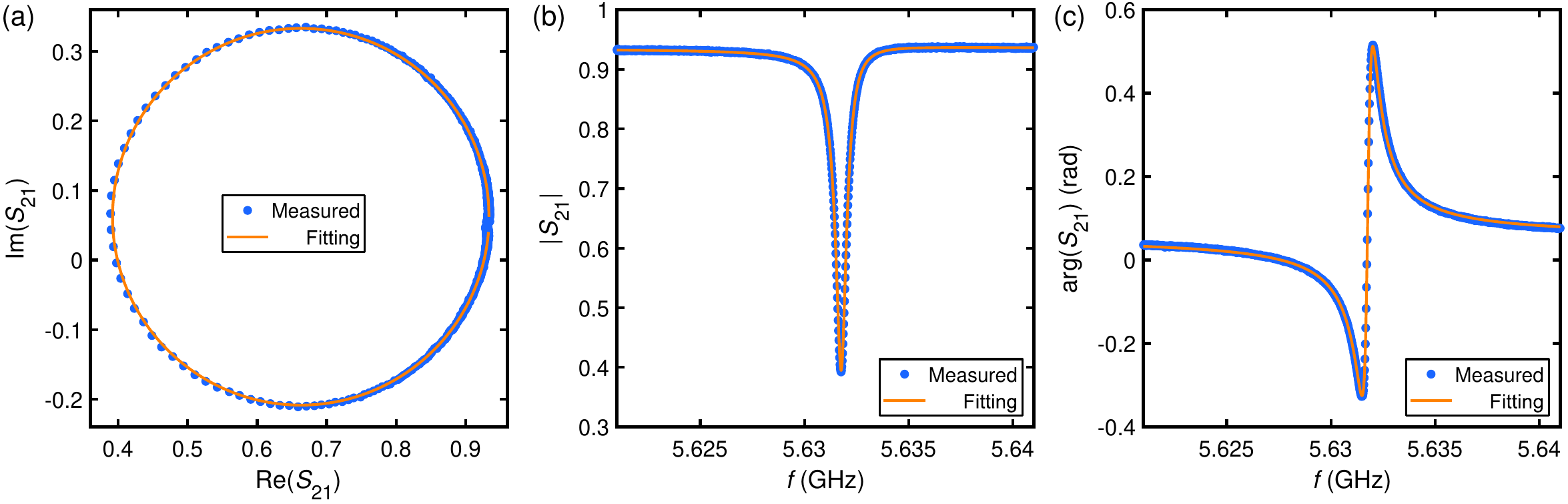}
     \caption{a) Notch-type scattering parameter circle for 25~nm thick CoSi\textsubscript{2} cavity. Frames b) and c) display the magnitude and phase response, respectively. A substantial mixing of the real and imaginary components of ideal resonance behavior is visible in the data. The fit curves are calculated using Eq.~(\ref{eq:notch}) with parameters $\omega_\mathrm{r}/2\pi =$\num{5.632e9}~s$^{-1}$, $Q_\mathrm{load} =$\num{7092}, $Q_\mathrm{cpl} =$\num{12236}, $\phi =$\num{-0.097}~rad, $a =$\num{0.9348}, $\alpha =$\num{0.2939} and $\tau =$\num{6.767e-12}~s.}
     \label{fig:resonance}
 \end{figure*}

 The impedance of a notch type of resonator including its environment leads to scattering parameter $S_{21}$ given by Eq.~(\ref{eq:notch}). The measured response curve is illustrated in Fig.~\ref{fig:resonance}b, in which the large suppression at resonance indicates that our device is nearly critically coupled.
 
Fig.~\ref{fig:resonance} displays fits of Eq.~(\ref{eq:notch}) to the measured data in a 25-nm-thick resonator. In both 10~nm and 105~nm thick resonators, the real and imaginary components of a Lorentzian response function are strongly mixed, while the mixing is nearly absent in the 25~nm thick cavities. Eq.~(\ref{eq:notch}) fits our data well in the linear response regime. When approaching the critical drive amplitude for the hysteretic Duffing regime, the fits deteriorate, and they become fully impossible in the hysteretic regime (part of the circle is missing). In addition to Duffing behavior, the 10~nm samples were found to display heating effects, which led to extra hysteresis due to frequency shifts caused by $L_\mathrm{K}(T)$. The analyzed results in this work are based on resonance shapes and fit close to that in Fig.~\ref{fig:resonance}.

\begin{table}[!htbp]
\vspace{10pt}
    \centering
    \begin{tabular}{cccccccc}
         \toprule
         & $f_\mathrm{dgn}$/GHz & $f_\mathrm{m}$/GHz & $L_\mathrm{line}$/mm & $L_\mathrm{K}/L_\mathrm{geo}$ & $Q_\mathrm{int}$ & $Q_\mathrm{cpl}$ & $Q_\mathrm{load}$\\
         \hline
         $f_1$ & 6.83 & 5.208 & 4.36 & 0.720 & 6930 & 20910 & 5200\\
         \hline
         $f_2$ & 7.16 & 5.447 & 4.15 & 0.728 & 19170 & 12160 & 7440\\
         \hline
         $f_3$ & 7.38 & 5.632 & 4.04 & 0.717 & 18500 & 12760 & 7550\\
         \hline
         $f_4$ & 7.49 & 5.782 & 3.91 & 0.678 & 28180 & 11880 & 8360\\
         \hline
         $f_5$ & 7.73 & 5.882 & 3.86 & 0.727 & 25560 & 10480 & 7430\\
         \bottomrule
    \end{tabular}
    \caption{Characteristics of the investigated 25-nm-thick CoSi\textsubscript{2} resonators with a nominal characteristic impedance of $Z_0=\sqrt{L/C}=50$~$\Omega$; actual $Z_0=65$~$\Omega$. The design frequency $f_\mathrm{dgn}$ is calculated from the length $L_{\mathrm{line}}$ using the theoretical speed of propagation. $f_\mathrm{m}$ denotes the measured resonance frequency which yields the kinetic inductance to geometric inductance ratio $L_\mathrm{K}/L_\mathrm{geo}$. $Q_\mathrm{int}$, $Q_\mathrm{cpl}$, and $Q_\mathrm{load}$ denote internal, coupling, and loaded quality factors, respectively. 
    }
    \label{tab:I}
\end{table}
Table~\ref{tab:I} displays the fitting results for 25-nm-thick microwave cavities. The design frequencies $f_{\mathrm{dgn}}$ reflect the physical length of the resonators and the speed of propagation in the coplanar transmission line given by $1/\sqrt{\ell c}$ where $\ell$ and $c$ are the inductance and capacitance per unit length, respectively. Frequency $f_\mathrm{m}$ denotes the measured resonance frequency, which yields the ratio of kinetic and geometric inductances as $ L_{\mathrm{K}} / L_{\mathrm{geo}} = \left(f_{\mathrm{dgn}}/f_{\mathrm{m}}\right)^2 -1 $. The division of loaded $Q$ factor into coupling dissipation rate $\propto 1/Q_{\mathrm{cpl}}$ and inherent dissipation rate $\propto 1/Q_{\mathrm{int}}$ arises from $S_{21}$ circle size and its orientation given by Eq.~(\ref{eq:notch}). The obtained kinetic inductance is consistently around $L_\mathrm{K}/L_{\mathrm{geo}} = 0.72$, which indicates pretty uniform quality for the CoSi\textsubscript{2} film. For the inherent average dissipation from all data points, we find $Q_{\mathrm{ave}}=19700$. However, if we were to neglect the first resonator with a somewhat unrealistic coupling factor, we would find $Q_{\mathrm{ave}}=22900$, not far from the full average. 

\begin{figure}[!htbp]
    \centering
    \includegraphics[width=\linewidth]{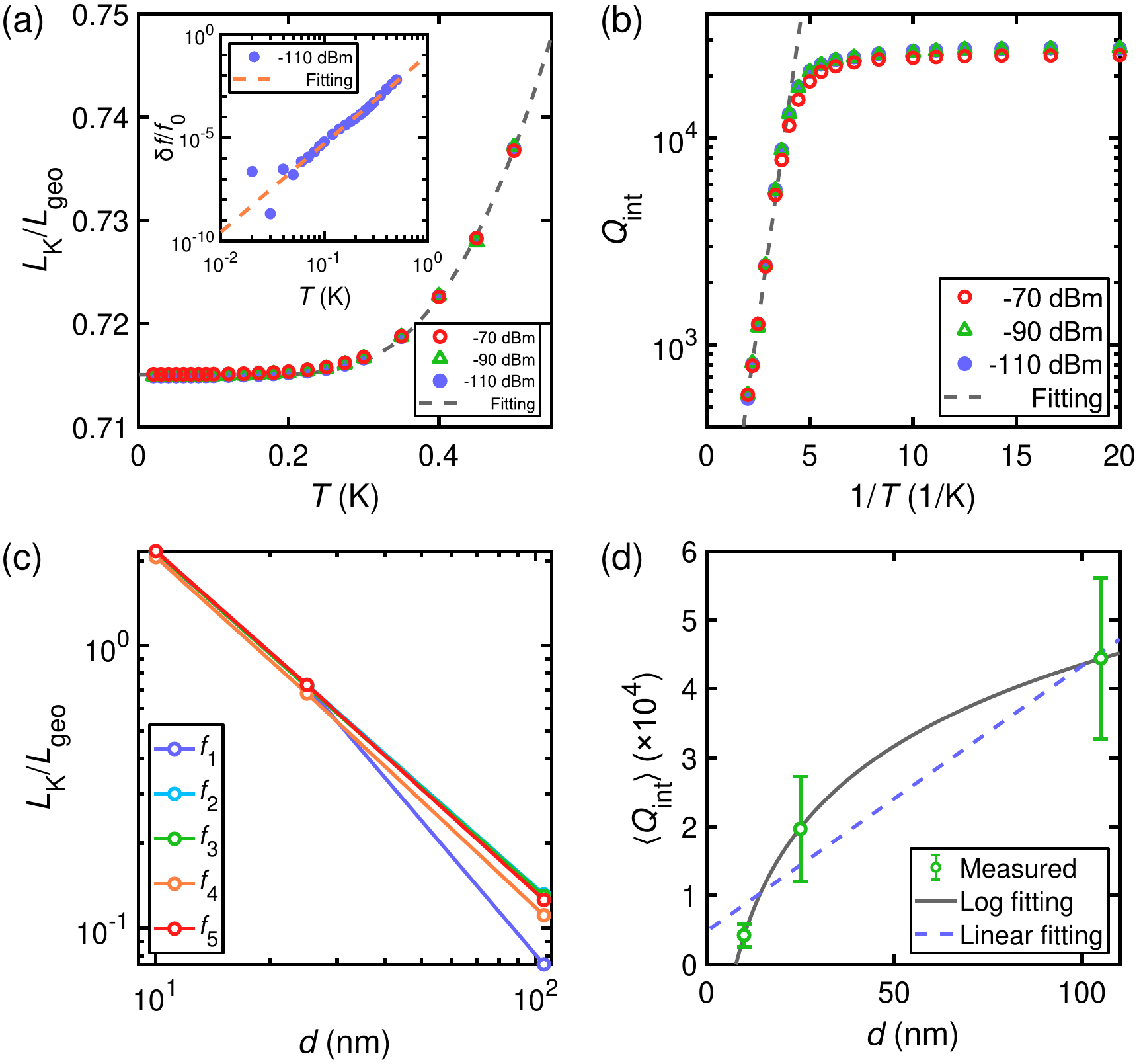}
        \caption{a) Kinetic inductance of 25~nm thick CoSi\textsubscript{2} cavity as a function of temperature. The fit illustrates $1/\Delta (T)$ dependence given by Eq.~(\ref{LK}) with $T_\mathrm{c}=1.07$~K. The inset shows the scaled resonance frequency shift $\delta f/f_0 = |f_\mathrm{m}(T)-f_\mathrm{m}(0)|/f_\mathrm{m}(0)$ as a function of temperature on log-log scale. The fitted dashed line yields $\delta f \propto T^{4.28}$.
        b) Simultaneously measured internal quality factor $Q_{\mathrm{int}}$ vs. temperature. The line depicts exponential dependence due to quasiparticle dissipation, $\exp(-\Delta/k_\mathrm{B} T)$ with constant $\Delta/k_\mathrm{B} = 1.63$~K.
        c) Kinetic inductance ($T=20$~mK) as a function of film thickness $d$ including all the measured cavities. The average data follow the $1/d$ dependence originating from approximate $R_\mathrm{n} \propto 1/d$ relation.  d) Average internal quality factor $ \langle Q_\mathrm{int}\rangle $ vs. film thickness $d$. The solid line illustrates a logarithmic fit while the dashed line denotes a linear fit.}
    \label{fig:Tdep}
\end{figure}

Fig.~\ref{fig:Tdep}a displays kinetic inductance $L_\mathrm{K}$ deduced from the resonance frequency of 25~nm thick CoSi\textsubscript{2} cavity as a function of temperature. Data obtained at three different measurement powers $P_\mathrm{m}$ are given (see the inset for $P_\mathrm{m}$ values). Already when $T > 0.1$~K, a distinct change in the temperature-dependent kinetic inductance $L_\mathrm{K}(T)$ is observed. The fit illustrates $1/\Delta (T)$ dependence given by Eq.~(\ref{LK}) using $L_{\mathrm{geo}}\simeq 1.4$~nH and $T_\mathrm{c} = 1.07$~K. The substantial deviation from the actual $T_\mathrm{c} = 1.31$~K may be an indication that the simple formula $L_\mathrm{K}(T) \propto R_\mathrm{n}/\Delta (T)$ is not fully accurate in our system. The inset displays the frequency change $\delta f$ relative to $T=0$ value $\delta f/f_\mathrm{0} = |f_\mathrm{m}(T)-f_\mathrm{m}(0)|/f_\mathrm{m}(0)$ vs. $T$ on a log-log scale. Our data display power law dependence $\propto T^{\alpha}$ with an exponent $\alpha = 4.3 \pm 0.2$. This finding differs from the standard behavior of metallic materials with similar sheet resistance \cite{Coumou2013}, and it calls either for non-BCS mechanisms, for example triplet type pairing \cite{Chiu2021}, or a suitable phonon density of states \cite{Eliashberg1991} for creation of elementary excitations.  

Fig.~\ref{fig:Tdep}b displays the internal quality factor $Q_\mathrm{int}$ extracted from the same data as used for Fig.~\ref{fig:Tdep}a (the inset indicates the values of $P_\mathrm{m}$). The data display universal exponential behavior at high $T$: the fitted dashed line depicts the exponential dependence due to quasiparticle dissipation, $Q_\mathrm{int} \propto R_{\mathrm{qp}} \propto \exp(\Delta/k_\mathrm{B} T)$ with constant $\Delta/k_\mathrm{B} = 1.63$~K. Using the BCS relation $\Delta/k_\mathrm{B} = 1.76 T_\mathrm{c}$, we obtain $T_\mathrm{c}=0.93$~K. This suppression of apparent $T_\mathrm{c}$ indicates either local reduction of the gap at some weakly ordered CoSi\textsubscript{2} spots or significant dissipation facilitated by the non-uniform gap in a triplet part of the CoSi\textsubscript{2} order parameter \cite{Chiu2021}.   

Kinetic inductance is found to follow scaling with $1/d$ as seen from the data in Fig.~\ref{fig:Tdep}c which depicts $L_\mathrm{K}$ values of all five resonators measured for each thickness. The largest thickness data having $d=105$~nm displays the biggest scatter owing to small kinetic inductance and to the insensitivity of the cavity properties to the kinetic inductance when $L_\mathrm{K}/L_{\mathrm{geo}} \ll 1$. 

Fig.~\ref{fig:Tdep}d displays the internal quality factor as a function of thickness. The data displays a strong, logarithmic-looking dependence on $d$ (see the solid curve), although the relative change is quite large for typical logarithmic dependence. 
For comparison, we have included a fit using linear thickness dependence (dashed line). As seen from the error bars ($=\sigma/\sqrt{N}$, where $\sigma$ is the standard deviation of $N=5$ cavities), the fluctuation between separate cavities is quite large, which makes it hard to distinguish reliably between these two fits.

\begin{center}
    \begin{figure}[]
        \centering
        \includegraphics[width=\linewidth]{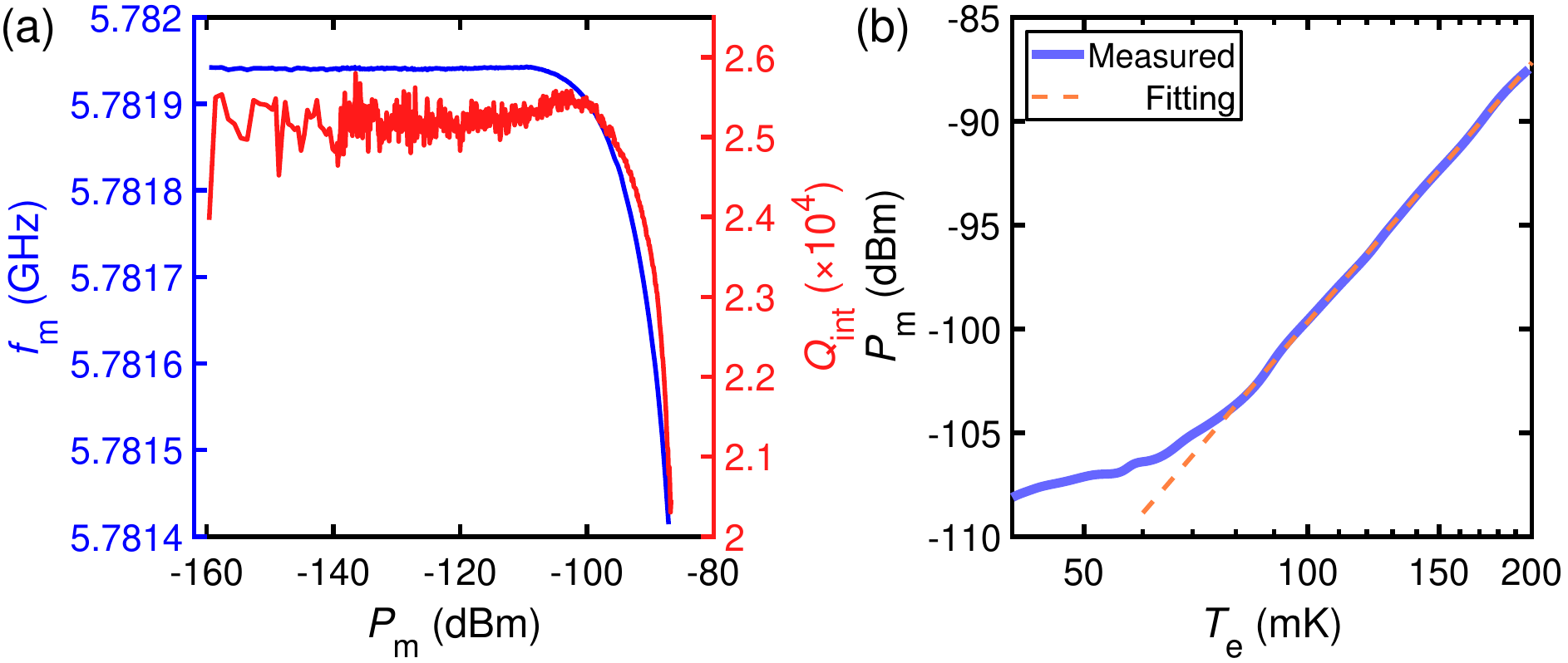}
            \caption{a) Power dependence of the cavity frequency $f_\mathrm{m}$ (upper curve, left scale) and the internal quality factor $Q_{\mathrm{int}}$ (lower curve, right scale). b) Electronic temperature $T_\mathrm{e}$ (deduced from the data for $f_\mathrm{m}$ in frame a) as a function carrier power. The solid curve is a fit for electron-phonon heat transfer using Eq.~(\ref{eq:heat}) with parameters $\gamma = 4.15$ and 
            $\rho_\mathrm{n}/\rho=1.0\times10^{-3}$.}
        \label{fig:response}
    \end{figure}
\end{center}

The dependence of the cavity resonance frequency $f_\mathrm{m}$ and the inherent quality factor $Q_\mathrm{int}$ on the drive power $P_\mathrm{m}$ (or equivalently, the number of photons in the cavity) is depicted in Fig.~\ref{fig:response}a. The data display a clear decrease in frequency and increase in dissipation with increasing drive power. We assign most of the increased dissipation to enhanced temperature with smaller quasiparticle resistance $R_{\mathrm{qp}}$. Thus, the measured $f_\mathrm{m}(P_\mathrm{m})$ and $Q_\mathrm{int}(P_\mathrm{m})$ can be employed for thermometric purposes via $f_\mathrm{m}(T)$ and $Q_\mathrm{int}(T)$ dependencies, respectively. The resolution of this kind of thermometry is clearly better using the shift in $f_\mathrm{m}$.

Fig.~\ref{fig:response}b shows the electronic temperature $T_\mathrm{e}$ obtained from the frequency shift in Fig.~\ref{fig:response}a and $f_\mathrm{m}(T)$ in the inset of Fig.~\ref{fig:Tdep}a, and depicts $T_\mathrm{e}$ as a function of microwave carrier power $P_\mathrm{m}$. By assuming that all of the power will be dissipated in the cavity \footnote{There is no galvanic contact facilitating electronic heat conduction away from the cavity.}, the dependence in Fig.~\ref{fig:response}b can be employed for estimation of electron-phonon (e-ph) coupling in this system. Similar estimation of e-ph coupling has been performed for example in suspended graphene \cite{Laitinen2014}. A direct power law fit yields for electron-phonon heat flow $\dot{Q} \propto T_\mathrm{e}^4$, but one should keep in mind that only the fraction of normal electrons carries the heat (see below). This power law exponent is close to that found in thermal experiments on niobium superconductor \cite{Feshchenko2017}.

Our data on $Q_{\mathrm{int}}$ vs. $P_\mathrm{m}$ covers also small powers (see Fig.~\ref{fig:response}a). The results showed only a \numrange[range-phrase = --,range-units = single]{1}{2}\% increase with number of photons in the range \numrange[range-phrase = --]{1}{1000} quanta in the cavity. Consequently, we believe that microwave dissipation due to TLSs is quite small in these CoSi\textsubscript{2} devices \cite{Wang2020}, which is consistent with the observed extremely limited disorder at the CoSi\textsubscript{2}/Si interfaces \cite{Chiu2017}. Most likely, the weak power dependence of $Q_\mathrm{int}(P_\mathrm{m})$ is due to a small number of TLSs combined with presence of an extra dissipation channel, e.g. due to vortex flow \cite{Clem1992,Song2009}. 

\section{Discussion}

Our results display quite strong dependence of internal quality factor $Q_{\mathrm{int}}$  on the thickness of the cavity material. According to Fig.~\ref{fig:Tdep}d, $Q_{\mathrm{int}} \propto \log(d)$ which is difficult to explain by geometrical changes in circuit parameters \footnote{We can approximately estimate the capacitance of the coplanar conductor as that of two parallel wires (bottom ground plane is far away), for which it is known that $C \propto 1/\log(2a/d)$ where $a$ is the separation of wires with diameter $d$. Modeling dissipation as $R$ in parallel to $C$, the effective $Q$ factor of the capacitance is given by $Q=\omega C R$, which yields $Q\propto \log (d)$ in the first order expansion when $d \ll a$. However, the change in $Q$ in this model would not be as large as observed in the experiment.}. One option could be that not all of the cobalt has formed CoSi\textsubscript{2} and there are magnetic cobalt moments present leading to active resistance in the proximity of the superconductor. Thus, CoSi\textsubscript{2} would have a boundary layer that is able to carry dissipative normal current. Owing to modified flux screening with increased thickness, this effective resistance could grow substantially and account for the strong increase seen in the experiment. However, this picture is contrary to the magnetization measurements of Ref.~\onlinecite{Chiu2021}, which indicate the absence of magnetic moments.

As seen in Fig.~\ref{fig:Tdep}d, the data can also be fit, within the scatter, with a linear thickness dependence $Q \propto d$. This linear behavior is suggestive for vortex pinning with the effective pinning strength growing linearly with the thickness. Owing to the large spread of surface roughness (see Fig.~\ref{fig:layout}c), the pinning behavior could vary quite a lot between our samples, which would explain most of the large scatter seen in our experiments. Our experiments were performed in Earth's magnetic field (\qty{\sim 50}{\micro T}) which is known to influence cavities made of Al (150~nm thick) and Re (50~nm thick) \cite{Song2009} that are close in characteristics in comparison with our 105-nm devices. Most likely, the spread in pinning potential strength increases with annealing time (thicker films), which could lead to larger dissipation at small $d$ as seen in the experiments. However, further experiments are needed to separate the effect of TLS-based dissipation from flux-flow dissipation in our devices.

For normal electrons in metals, the heat transfer by electron-phonon coupling can be written as \cite{Sergeev1996,Timofeev2009,Heikkila2018,Nikolic2020}
\be \label{eq:heat}
\dot{Q} \simeq \Sigma V_\mathrm{n}(T_\mathrm{e}^{\gamma} - T_\mathrm{ph}^{\gamma}),
\ee
where parameter $\Sigma \simeq 10^9$~Wm\textsuperscript{-3}K\textsuperscript{-5} for bulk metals with $\gamma =5$ \cite{Giazotto2006}, $V_\mathrm{n}=(\rho_\mathrm{n}/\rho)V$ is the effective volume of heat-carrying electrons (fractional density $\rho_\mathrm{n}/\rho$) in total volume $V$, while $T_\mathrm{e}$ and $T_\mathrm{ph}$ are the electron and phonon temperatures, respectively. In disordered, thin conductors, the exponent $\gamma$ depends on the disorder and phonon dimensionality, resulting in $\gamma =$~\numrange[range-phrase = --]{4}{6} \cite{Roukes1985,Sergeev2000,Karvonen2005}. 

We obtained exponent $\gamma =4.15$ by fitting Eq.~(\ref{eq:heat}) to the measured $\dot{Q}(T_\mathrm{e})$ in Fig.~\ref{fig:response}b using 
$\rho_\mathrm{n}/\rho=1.0 \times 10^{-3}$, $V=1.25\times10^{-15}$~m$^3$ for the cavity volume, and $\Sigma \simeq 10^9$~Wm\textsuperscript{-3}K\textsuperscript{-4.15}. The employed prefactor $\Sigma$ corresponds to regular results on metals around 1~K temperature (see, e.g. Ref.~\onlinecite{Giazotto2006}). The residual normal electron fraction corresponds to $T=0.22$~K ($T=0.33$~K) if considered as thermally excited quasiparticles in the sample using $\Delta/k_\mathrm{B}=1.63$~K ($\Delta/k_\mathrm{B}=1.76 T_\mathrm{c}$), which is clearly higher than the lower range of temperatures in \qtyrange[range-phrase = --,range-units = single]{100}{200}{mK}. This discrepancy in comparison with typical excess quasiparticle density \cite{Visser2012,Visser2014} may indicate the presence of anisotropic triplet correlations in CoSi\textsubscript{2}\cite{Chiu2021} or additional relaxation channel from CoSi\textsubscript{2} electrons to substrate phonons \cite{Elo2017}.

We also estimated the e-ph coupling from 10~nm and 105~nm samples. However, the results were much more unreliable compared with 25~nm data. The data on 10~nm cavities were hampered by the small $Q$ factor and an early onset of non-linear behavior, while the resolution of kinetic inductance thermometry in 105~nm devices was bad from the beginning. Thus, we are not able to conclude anything about the universality of the exponent $\gamma=4.15$ as a function of thickness, which would illuminate the issue of mixed-dimensionality phonons in thin films \cite{Anghel2019}.  

To conclude, our study shows that CoSi\textsubscript{2} is a promising material for quantum circuits due to its high-quality, low-temperature superconducting properties, and excellent lattice match to silicon. We demonstrate that thin CoSi\textsubscript{2} films possess a substantial kinetic inductance, which displays a non-BCS power-law dependence $\delta L_\mathrm{K} \propto T^{4.3 \pm 0.2}$ at the lowest temperatures.  
Our investigation also addresses the potential issue of enhanced dissipation at microwave frequencies caused by two-level tunneling states, as well as the efficiency of thermalization of charge carriers at large drives. Our findings suggest that CoSi\textsubscript{2} is a valuable addition to the toolkit of materials for quantum circuit fabrication, with applications in ultra-low temperature sensing, single photon detection, and parametric amplification.

	\section*{Acknowledgments}
 We are grateful to Tero Heikkil\"a, Alexander Zyuzin, Teun Klapwijk, Manohar Kumar, P\"aivi T\"orm\"a, and Shao-Pin Chiu for fruitful discussions, and Pei-Ling Wu and Shouray Kumar Sahu for experimental assistance. This work was supported by the Academy of Finland (AF) projects 341913 (EFT), and 312295 \& 352926 (CoE, Quantum Technology Finland), as well as Taiwan-Finland AF mobility grant 341884 (S.~S.~Yeh). The research leading to these results has received funding from the European Union’s Horizon 2020 Research and Innovation Programme, under Grant Agreement No.~824109 (EMP). Our work was also supported by funding from Jane and Aatos Erkko Foundation and Keele Foundation (SuperC project). The experimental work benefited from the Aalto University OtaNano/LTL infrastructure. J.~J.~Lin acknowledges the support by the National Science and Technology Council (NSTC) of Taiwan through Grant Nos.~110-2112-M-A49-015 and 111-2119-M-007-005. 
 S.~S.~Yeh is grateful to the support by NSTC of Taiwan through Grant No.~110-2112-M-A49-033-MY3 and the support by the Taiwan Ministry of Education through the Higher Education Sprout Project of the NYCU.

\section*{DATA AVAILABILITY STATEMENT}
 The data that support the findings of this study are available from the corresponding author upon reasonablebrequest.
 




 %

	\appendix

\end{document}